\documentclass{ws-ijmpd}
\usepackage[super,compress]{cite}
\begin{document}

\markboth{Andrei G. Lebed}
{Inequivalence between active and passive gravitational masses and energy}

%
\catchline{}{}{}{}{}
%

\title{{INEQUIVALENCE BETWEEN GRAVITATIONAL MASS AND ENERGY DUE TO QUANTUM EFFECTS AT MICROSCOPIC AND MACROSCOPIC LEVELS \\
}}

\author{ANDREI G. LEBED}

\address{Department of Physics, University of Arizona, 1118 E. 4th Street,\\
Tucson, Arizona 85721, USA and\\
L.D. Landau Institute for Theoretical Physics, RAS, 2 Kosygina Street,\\
Moscow, 117334 Russia\\
lebed@physics.arizona.edu}

\maketitle

\begin{history}
\received{Day Month Year}
\revised{Day Month Year}
\end{history}

\begin{abstract}
We review recent theoretical results, demonstrating breakdown of
the equivalence between active and passive gravitational masses
and energy due to quantum effects in General Relativity. In
particular, we discuss the simplest composite quantum body - a
hydrogen atom - and define its gravitational masses operators.
Using Gedanken experiment, we show that the famous Einstein's
equation, $E =mc^2$, is broken with small probability for passive
gravitational mass of the atom. It is important that the
expectation values of both active and passive gravitational masses
satisfy the above mentioned equation for stationary quantum
states. Nevertheless, we stress that, for quantum superpositions
of stationary states in a hydrogen atom, where the expectation
values of energy are constant, the expectation values of the
masses oscillate in time and, thus, break the Einstein's equation.
We briefly discuss experimental possibility to observe the above-mentioned time-dependent oscillations. In this review, we also improve several drawbacks of the original pioneering works.   
\end{abstract}

\keywords{Equivalence principle; Mass-energy equivalence; Quantum
gravity.}

\ccode{PACS numbers: 04.60.-m, 04.80.Cc}


\section{Introduction}
It is well known that creation of the Quantum Gravitational theory
is the most important step in development of the so-called Theory
of Everything. This problem seems to be extremely difficult.
This is partially due to the fact that the foundations of Quantum
Mechanics and General Relativity are very different and partially
due to the absence of the corresponding experimental data. So far,
quantum effects have been directly experimentally observed only in
the Newtonian variant of gravitation [1,2]. On the other hand,
such important quantum phenomenon in General Relativity as the
Hawking radiation [3] is still very far from its direct
experimental discovery. In this complicated situation, we have
recently suggested two novel phenomena [4-10], which show that
active and passive gravitational masses of a composite quantum
body are not equivalent to its energy due to some quantum effects.
Moreover, we have also discussed [9] possible experimental way to observe
one the above mentioned phenomena - time-dependent oscillations of
active gravitational mass for quantum superpositions of stationary
states. We have also suggested several methods to derive quantum
mechanical operators of passive and active gravitational masses in
a hydrogen atom: semi-quantitative ones [4-9] and direct method to
derive the operators from the Dirac equation in a curved spacetime
of General Relativity [10].

We stress that a notion of passive gravitational mass of a
composite body is not trivial even in classical physics. Let us
consider a classical model of a hydrogen atom. As mentioned by
Nordtvedt [11] and Carlip [12], external weak gravitational field
is coupled with the following combination: $m_e + m_p + (3K +
2P)/c^2$, where c is the velocity of light, $m_e$ and $m_p$ are
the bare electron and proton masses, $K$ and $P$ are electron
kinetic and potential energies. Nevertheless, due to the classical
virial theorem averaged over time combination $<2K + P>_t =0$ and,
therefore, averaged over time electron passive gravitational mass,
$<m^p_e>_t$, satisfies the famous Einstein's equation,
\begin{equation}
<m^p_e>_t = m_e + \biggl< \frac{K + P}{c^2} \biggl>_t + \biggl<
\frac{2K + P}{c^2} \biggl>_t = m_e + \biggl< \frac{K + P}{c^2}
\biggl>_t = \frac{E}{c^2},
\end{equation}
where $E$ is the total electron energy. On this basis, in
Refs.[11,12], the conclusion about the equivalence between
averaged over time passive gravitational mass and energy was made.
As to active gravitational mass, it has been shown that it is also
non-trivial notion for a composite body even in classical case and
is related to the following interesting paradox. If we
apply the so-called Tolman's formula for active gravitational mass
[13],
\begin{equation}
m_{ph}^a = \frac{1}{c^2} \int [T_0^0({\bf r}) - T_1^1({\bf r}) -
T_2^2({\bf r}) -T_3^3({\bf r})] \ d^3 r \ ,
\end{equation}
to a free photon with energy $E$, we obtain $m_{ph}^a = 2E/c^2$
(i.e., two times bigger value than the expected one)[14]. Let us
now consider the photon in a box with mirror walls (i.e., a
composite body at rest). Then, as shown by Misner and Putnam [14],
the Einstein's equation, $m^a_{ph}=E/c^2$, restores, if we take
into account negative contribution to active gravitational mass
from stress in the box walls. So, in the example above, both
kinetic and potential energies make contributions to active
gravitational mass and the Einstein's equation is restored only
after averaging over time [14].

\section{Goal}

The goal of our review is to describe in detail the recent results
[4-10], related to breakdown of the equivalence between energy and
active and passive gravitational masses due to quantum effects.
Note that our conclusions are applicable to any composite quantum
body, although below we consider the simplest example of such a
body - a hydrogen atom in the Earth's gravitational field. In Section
3, we consider electron active gravitational mass in the atom and
show that its expectation value is equivalent to energy for
stationary electron quantum states. On the other hand, we
demonstrate that this equivalence is broken for quantum
superpositions of stationary states. In particular, we show that
the expectation value of active gravitational mass exhibits
time-depended oscillations even in superposition of the states,
where the expectation value of energy is constant [6,9]. It is important that this corresponds to breakdown of the equivalence between
active gravitational mass and energy at macroscopic level. We also
discuss in brief idealized experiment, which can discover the above
mentioned breakdown. In Section 4, we concentrate on study of
electron passive gravitational mass in a hydrogen atom. We derive
the corresponding mass operator using four different ways,
including direct consideration of the Dirac equation in a curved
spacetime of General Relativity. We discuss Gedanken experiment,
which shows inequivalence of electron passive gravitational mass
and energy at a microscopic level [4-6,10]. As to a macroscopic level, the expectation value of electron passive gravitational mass is shown
to be equivalent to energy only for stationary quantum states.
Nevertheless, for quantum superpositions of stationary states, the
equivalence between the expectation values of passive gravitational
mass and energy is shown to be broken due to time-dependent
oscillations of the expectation values of the mass [4,6].

\section{Active gravitational mass [6,9]}
In this Section, we derive expression for electron active
gravitation mass in a classical model of a hydrogen atom in the
so-called post-Newtonian approximation of General Relativity [15].
Then, we quantize it and use the so-called semi-classical
Einstein's gravitational field equation [16].

\subsection{Active gravitational mass in classical physics}
Here, we determine electron active gravitational mass in a
classical model of a hydrogen atom, which takes into account
electron kinetic and potential energies. More specifically, we
consider a particle with small bare mass $m_e$, moving in the
Coulomb electrostatic field of a heavy particle with bare mass
$m_p \gg m_e$. Our task is to find gravitational potential at large
distance from the atom, $R \gg r_B$, where $r_B$ is the the
so-called Bohr radius (i.e., effective "size" of a hydrogen atom).
Bellow, we use the so-called weak field gravitational theory
[13,15], where the post-Newtonian gravitational potential can be
represented as [6,9]

\begin{equation}
\phi(R,t)=-G \frac{m_p + m_e}{R}- G \int \frac{\Delta
T^{kin}_{\alpha \alpha}(t,{\bf r})+
 \Delta T^{pot}_{\alpha
\alpha}(t,{\bf r})}{c^2R} d^3 {\bf r} ,
\end{equation}
where $\Delta T^{kin}_{\alpha \beta}(t,{\bf r})$ and $\Delta
T^{pot}_{\alpha \beta}(t,{\bf r})$ are contributions to
stress-energy tensor density, $T_{\alpha \beta}(t, {\bf r})$, due
to kinetic and the Coulomb potential energies, respectively. We
point out that, in Eq.(3), we disregard all retardation effects.
Thus, in the above-discussed approximation, electron active
gravitational mass equals to
\begin{equation}
m^a_e = m_e + \frac{1}{c^2} \int [\Delta T^{kin}_{\alpha
\alpha}(t,{\bf r}) + \Delta T^{pot}_{\alpha \alpha}(t,{\bf r})]
d^3{\bf r}.
\end{equation}

Let us calculate $\Delta T^{kin}_{\alpha \alpha}(t, {\bf r})$,
using the standard expression for stress-energy tensor density of
a moving relativistic point mass [13,15]:
\begin{equation}
T^{\alpha \beta}({\bf r},t) = \frac{m_e v^{\alpha}(t)
v^{\beta}(t)}{\sqrt{1-v^2(t)/c^2}} \ \delta^3[{\bf r}-{\bf
r}_e(t)],
\end{equation}
where $v^{\alpha}$ is a four-velocity, $\delta^3(...)$ is the
three dimensional Dirac $\delta$-function, and ${\bf r}_e(t)$ is a
three dimensional electron trajectory.

From Eqs.(4),(5), it directly follows that
\begin{equation}
\Delta T^{kin}_{\alpha \alpha}(t) = \int \Delta T^{kin}_{\alpha
\alpha}(t,{\bf r}) d^3{\bf r} = \frac{m_e [c^2
+v^2(t)]}{\sqrt{1-v^2(t)/c^2}} -m_ec^2.
\end{equation}
Note that, although calculations of the contribution from potential
energy to stress energy are more complicated, they are straightforward
and can be done by using the standard formula for stress energy
tensor of electromagnetic field [13],
\begin{equation}
T_{em}^{\mu \nu} = \frac{1}{4 \pi} [F^{\mu \alpha} F^{\nu}_{\
\alpha} - \frac{1}{4} \eta^{\mu \nu} F_{\alpha \beta} F^{\alpha
\beta}],
\end{equation}
where $\eta_{\alpha \beta}$ is the Minkowski metric tensor,
$F^{\alpha \beta}$ is the so-called tensor of electromagnetic
field [13]. In this review, we use approximation, where we do not
take into account magnetic field and keep only the Coulomb
electrostatic field. In this approximation, we can simplify Eq.(7)
and obtain from it the following expression:
\begin{equation}
\Delta T^{pot}_{\alpha \alpha} (t) = \int \Delta T^{pot}_{\alpha
\alpha}(t,{\bf r}) d^3{\bf r} = -2\frac{e^2}{r(t)},
\end{equation}
where $e$ is the electron charge.
As directly follows from Eqs.(6),(8), electron active
gravitational mass can be represented in the following way:
\begin{equation}
m^a_e = \biggl[\frac{m_e c^2}{(1 -v^2/c^2)^{1/2}} - \frac{e^2}{r}
\biggl]/c^2 + \biggl[\frac{m_e v^2}{(1
-v^2/c^2)^{1/2}}-\frac{e^2}{r}\biggl]/c^2.
\end{equation}
We note that the first term in Eq.(9) is the expected one. Indeed,
it is the total energy contribution to the mass, whereas the
second term is the so-called relativistic virial one [17]. It is
important that it depends on time. Therefore, in classical
physics, active gravitational mass of a composite body depends on
time too. Nevertheless, in this situation, it is possible to
introduce averaged over time electron active gravitational mass.
This procedure results in the expected equivalence between
averaged over time active gravitational mass and energy:
\begin{equation}
<m^a_e>_t = \biggl<\frac{m_e c^2}{(1 -v^2/c^2)^{1/2}} -
\frac{e^2}{r} \biggl>_t/c^2 + \biggl<\frac{m_e v^2}{(1
-v^2/c^2)^{1/2}}-\frac{e^2}{r}\biggl>_t/c^2 = m_e + E/c^2 .
\end{equation}
We point out that, in Eq.(10), the averaged over time virial term
is zero due to the classical virial theorem. It is easy to show
that for non-relativistic case our Eqs.(9),(10) can be simplified
to the results of Refs.[11,12]:
\begin{equation}
m^a_e = m_e + \biggl(\frac{m_e v^2}{2} - \frac{e^2}{r} \biggl)/c^2
+ \biggl(2 \frac{m_e v^2}{2}-\frac{e^2}{r}\biggl)/c^2
\end{equation}
and
\begin{equation}
<m^a_e>_t = m_e + \biggl<\frac{m_e v^2}{2} - \frac{e^2}{r}
\biggl>_t/c^2 + \biggl<2 \frac{m_e
v^2}{2}-\frac{e^2}{r}\biggl>_t/c^2 = m_e + E/c^2.
\end{equation}

\subsection{Active gravitational mass in quantum physics [6,9]}

In this Subsection, we consider the so-called semiclassical theory
of gravity [16], where, in the Einstein's field equation,
gravitational field is not quantized but the matter is quantized:
\begin{equation}
R_{\mu \nu} - \frac{1}{2}R g_{\mu \nu} = \frac{8 \pi G}{c^4}
\bigl<\hat T_{\mu \nu} \bigl> .
\end{equation}
Here, $<\hat T_{\mu \nu}>$ is the expectation value of quantum
operator, corresponding to the stress-energy tensor. To make use
of Eq.(13), we have to rewrite Eq.(11) for electron active
gravitational mass using momentum, instead of velocity. Then, we
can quantize the obtained result:
\begin{equation}
\hat m^a_e = m_e +\biggl(\frac{{\bf \hat
p}^2}{2m_e}-\frac{e^2}{r}\biggl)/c^2 + \biggl(2\frac{{\bf \hat
p}^2}{2m_e}-\frac{e^2}{r}\biggl)/c^2.
\end{equation}
Note that Eq.(14) represents electron active gravitational mass
operator. As directly follows from it, the expectation value of
electron active gravitational mass can be written as
\begin{equation}
<\hat m^a_e> = m_e +\biggl< \frac{{\bf \hat
p}^2}{2m_e}-\frac{e^2}{r}\biggl>/c^2 + \biggl<2\frac{{\bf \hat
p}^2}{2m_e}-\frac{e^2}{r}\biggl>/c^2,
\end{equation}
where third term is the virial one.

\subsubsection{Equivalence of the expectation values [6,9]}

Now, we consider a macroscopic ensemble of hydrogen atoms with
each of them being in the n-th energy level. For such ensemble,
the expectation value of the mass (15) is
\begin{equation}
<\hat m^a_e> = m_e + \frac{E_n}{c^2}.
\end{equation}
In Eqs(15),(16), we take into account that the expectation value
of the virial term is equal to zero in stationary quantum states
due to the quantum virial theorem [17]. Thus, we can make the
following important conclusion: in stationary quantum states,
active gravitational mass of a composite quantum body is
equivalent to its energy at a macroscopic level [6,9].

\subsubsection{Inequivalence between active gravitational mass and
energy at a macroscopic level [6,9]}

Below, we introduce the simplest quantum superposition of the
following stationary states in a hydrogen atom,
\begin{equation}
\Psi (r,t) = \frac{1}{\sqrt{2}} \bigl[ \Psi_1(r) \exp(-iE_1t)
+ \Psi_2(r) \exp(-iE_2t) \bigl],
\end{equation}
where $\Psi_1(r)$ and $\Psi_2(r)$ are the normalized wave functions of the
ground state (1S) and first excited state (2S), respectively. It
is easy to show that the superposition (17) corresponds to the
following constant expectation value of energy:
\begin{equation}
<E> = \frac{E_1+E_2}{2}.
\end{equation}
Nevertheless, as seen from Eq.(15), the expectation value of
electron active gravitational mass operator for the wave function (17)
is not constant and exhibits time-dependent oscillations:
\begin{equation}
<\hat m^a_e> = m_e + \frac{E_1+E_2}{2 c^2} + \frac{V_{1,2}}{c^2}
\cos \biggl[ \frac{(E_1-E_2)t}{\hbar} \biggl],
\end{equation}
where $V_{1,2}$ is matrix element of the virial operator,
\begin{equation}
V_{1,2} = \int \Psi_1(r) \ \biggl(2\frac{{\bf \hat
p}^2}{2m_e}-\frac{e^2}{r}\biggl) \ \Psi_2(r) \ d^3{\bf r} \ ,
\end{equation}
between the above-mentioned two stationary quantum states. It is
important that the oscillations (19),(20) directly demonstrate
breakdown of the equivalence between the expectation values of
active gravitational mass and energy for quantum superpositions of
stationary states [6,9]. We pay attention to the fact that such quantum
time-dependent oscillations are very general and are not
restricted by the case of a hydrogen atom. They are of a pure
quantum origin and do not have classical analogs. To make the
situation trivial, we can define the averaged over time
expectation value of electron active gravitational mass. We stress that, although the latter mass obeys the Einstein's equation,
\begin{equation}
<<\hat m^a_e>>_t = m_e + \frac{E_1+E_2}{2 c^2} = \biggl<
\frac{E}{c^2} \biggl>,
\end{equation}
the expectation values of active gravitational mass and energy are
shown by us to be inequivalent to each other for quantum
superpositions of stationary state.

\subsection{Suggested experiment}
In this short Subsection, we suggest an idealized experiment,
which allows to observe quantum time-dependent oscillations of
the expectation values of active gravitational mass (19). In
principle, it is possible to create a macroscopic ensemble of the
coherent quantum superpositions of electron stationary states in
some gas with high density. It is important that these
superpositions have to be characterized by the feature that each
molecule has the same phase difference between two wave function
components, $\tilde \Psi_1(r)$ and $\tilde \Psi_2(r)$. In this
case, the macroscopic ensemble of the molecules generates
gravitational field, which oscillates in time similar to Eq.(19),
which, in principle, can be measured. It is important to use such
geometrical distributions of the molecules and a test body, where oscillations (19) are "in phase" and do not cancel each other.

\section{Passive gravitational mass [4-8,10]}
In the beginning of this Section, we suggest several methods to
obtain expression for electron passive gravitational mass operator
in a hydrogen atom. Then, using the obtained expression, we
establish the equivalence between the expectation value of
electron passive gravitational mass and its energy for stationary
quantum states. For quantum superpositions of stationary states,
we obtain breakdown of the equivalence between the corresponding
expectation values due to quantum oscillations of the expectation
values of electron passive gravitational mass. The latter indicates
breakdown of the equivalence between the mass and energy at a
macroscopic level. In the end of this Section, we establish
breakdown of the equivalence between passive gravitational mass
and energy at a microscopic level (i.e., for an individual
measurement of electron passive gravitational mass in the atom).

\subsection{Lagrangian approach [4,5,7]}
In this Subsection, we derive expression for electron passive
gravitational mass operator by means of two methods, which make
use the Lagrangian approach.

\subsubsection{Passive gravitational mass in classical physics}

Below, we derive the Lagrangian and Hamiltonian of a classical model
of a hydrogen atom in the Earth's gravitational field, taking into
account couplings of electron kinetic and the Coulomb potential
energies with the gravitational field. We use the so-called
post-Newtonian approximation. In other words, we keep only terms
of the order of $1/c^2$ and disregard magnetic force as well as
radiations of both electromagnetic and gravitational waves. Here,
we also disregard all tidal and spin-dependent effects, which are
extremely small near the Earth. In other words, we write the
interval in the Earth gravitational field using the so-called weak
field approximation [13,15]:
\begin{equation}
d s^2 = -\biggl(1 + 2 \frac{\phi}{c^2} \biggl)(cdt)^2 + \biggl(1 -
2 \frac{\phi}{c^2} \biggl) (dx^2 +dy^2+dz^2 ), \ \phi = -
\frac{GM}{R} ,
\end{equation}
where $G$ is the gravitational
constant, $M$ is the Earth mass, $R$ is a distance between center
of the Earth and center of mass of a hydrogen atom (i.e., proton).
We point out that to calculate the Lagrangian (and later - the
Hamiltonian) in a linear with respect to the small parameter,
$|\phi(R)| / c^2 \ll 1$, approximation, we do not need to keep the
terms of the order of $[\phi(R)/c^2]^2$ in metric (22). This is in
a contrast to classical perihelion orbit procession calculations
[15].

As usual, for metric (22), it is possible to define the local proper
spacetime coordinates, where the metric has the Minkowski form,
\begin{equation}
x'=\biggl(1-\frac{\phi}{c^2} \biggl) x, \ y'=
\biggl(1-\frac{\phi}{c^2} \biggl) y, z'=\biggl(1-\frac{\phi}{c^2}
\biggl) z , \ t'= \biggl(1+\frac{\phi}{c^2} \biggl) t.
\end{equation}
If we disregard all tidal effects, in the above-mentioned coordinates,
the Lagrangian and action of a classical model of a hydrogen atom have
the following standard forms:
\begin{equation}
L' = -m_p c^2 -m_e c^2 + \frac{1}{2} m_e ({\bf v'})^2 +
\frac{e^2}{r'} \ , \ \ \ S' = \int L' dt' ,
\end{equation}
where ${\bf v'}$ is electron velocity and $r'$ is a distance
between electron and proton. [Note that, in our calculations, we
use the inequality $m_p \gg m_e$. In other words, we disregard
proton kinetic energy in the Lagrangian (24) and consider its
position as a position of center of mass of a hydrogen atom, which
is fixed by some force of a non-gravitational origin]. It is easy
to show that the Lagrangian (24) can be written in the global
coordinates $(x,y,z,t)$ as
\begin{equation}
L = -m_p c^2 -m_e c^2 +  \frac{1}{2}m_e{\bf v}^2+\frac{e^2}{r} - m_p \phi - m_e \phi - \biggl( 3m_e\frac{{\bf v}^2}{2}-2\frac{e^2}{r} \biggl) \frac{\phi}{c^2} .
\end{equation}

Now we calculate the Hamiltonian, which corresponds to the Lagrangian
(25), using the standard procedure, $H({\bf p},{\bf r})={\bf
p}{\bf v}-L({\bf v},{\bf r})$, where ${\bf p}= \partial L({\bf
v},{\bf r})/\partial {\bf v}$. As a result, we obtain:
\begin{equation}
H = m_p c^2 + m_e c^2 + \frac{{\bf p}^2}{2m_e}-\frac{e^2}{r} + m_p
\phi + m_e \phi + \biggl( 3 \frac{{\bf p}^2}{2 m_e}
-2\frac{e^2}{r} \biggl) \frac{\phi}{c^2},
\end{equation}
where canonical momentum in a gravitational field is ${\bf p}=m_e
{\bf v}(1-3\phi/c^2)$. We recall that, in this review, we
disregard all tidal effects. In particular, this means that we do
not differentiate the gravitational potential (22) with respect to
relative electron coordinates, ${\bf r}$ and ${\bf r'}$, which
correspond to electron position in center of mass coordinate
system. It is easy to demonstrate that this means that we consider
a hydrogen atom as a point-like body and disregard all effects of
the relative order of $r_B/R_0 \sim 10^{-17}$, where $R_0$ is the
Earth's radius. Let us reproduce the results of Refs.[11,12],
using the Hamiltonian formalism. Indeed, from the Hamiltonian
(26), we can define averaged over time electron passive
gravitational mass, $<m^p_e>_t$, as its average
 weight in the weak gravitational field (22). As a result, we obtain:
\begin{equation}
<m^p_e>_t = m_e  + \biggl<  \frac{{\bf p}^2}{2 m_e}-
\frac{e^2}{r}\biggl>_t \frac{1}{c^2} +\biggl< 2 \frac{{\bf p}^2}{2
m_e}-\frac{e^2}{r}\biggl>_t \frac{1}{c^2} =  m_e + \frac{E}{c^2} \
,
\end{equation}
where $E= {\bf p}^2/2 m_e - e^2/r$ is electron energy. We pay
attention that the averaged over time third term in Eq.(27) equals to
zero due to the classical virial theorem. Therefore, we can conclude that in classical
composite body passive gravitational mass, averaged over time, is
equivalent to the body energy, taken in the absence of
gravitational field [11,12].

\subsubsection{More general Lagrangian [11]}

Now, let us consider the Lagrangian of a three body system: a
hydrogen atom and the Earth in inertial coordinate system, related
to the center of mass (i.e., the Earth). In this case, we can make
use of the results of Ref.[11], where the corresponding n-body
Lagrangian is calculated as a sum of the following four terms:
\begin{equation}
L = L_{kin} + L_{em} + L_{G} + L_{e,G},
\end{equation}
where $L_{kin}$, $L_{em}$, $L_G$, and $L_{e,G}$ are kinetic,
electromagnetic, gravitational and electric-gravitational parts of
the Lagrangian, respectively. We recall that, in our
approximation, we keep in the Lagrangian and Hamiltonian only
terms of the order of $(v/c)^2$ and $|\phi|/c^2$ as well as keep
only classical kinetic and the Coulomb electrostatic potential energies
couplings with external gravitational field. It is possible to
show that, in our case, different contributions to the Lagrangian
(28) can be simplified:

\begin{equation}
L_{kin} + L_{em} = - Mc^2 - m_pc^2 - m_ec^2 +
m_e\frac{{\bf v}^2}{2} + \frac{e^2}{r} \ ,
\end{equation}

\begin{equation}
L_{G} = G \frac{m_p M}{R} + G \frac{m_e M}{R} + \frac{3}{2} G
\frac{m_e M}{R} \frac{{\bf v^2}}{c^2} \ ,
\end{equation}

\begin{equation}
L_{e,G} = - 2 G \frac{M}{Rc^2} \frac{e^2}{r} \ ,
\end{equation}
where, as usual, we use the inequality $m_p \gg m_e$.

If we keep only those terms in the Lagrangian, which are related
to electron motion (proton is supposed to be supported by some
non-gravitational force in the gravitational field), then we can
write the Lagrangian (28)-(31) in the following familiar form:
\begin{equation}
L = m_e \frac{{\bf v}^2}{2} + \frac{e^2}{r} - \frac{\phi(R)}{c^2}
\biggl[ m_e + 3 m_e \frac{{\bf v}^2}{2} - 2 \frac{e^2}{r} \biggl]
\ , \ \ \ \phi(R) = - G \frac{M}{R}.
\end{equation}
[Note that, as usual, we disregard the difference between electron
bare mass and the so-called reduced mass, which are almost equal
under the condition $m_p \gg m_e$.] It is easy to show that the
corresponding electron Hamiltonian is
\begin{equation}
H= \frac{{\bf p}^2}{2m_e} -\frac{e^2}{r} + \frac{\phi(R)}{c^2}
\biggl[ m_e + 3 \frac{{\bf p}^2}{2m_e} - 2 \frac{e^2}{r} \biggl] \ .
\end{equation}
It is important that Eqs.(32),(33) exactly coincide with electron
parts of the Lagrangian (25) and Hamiltonian (26), obtained by us
in the previous Subsubsection.

\subsubsection{Passive gravitational mass in quantum physics [4-6]}

Let us consider quantum problem about a hydrogen atom in the external
gravitational potential (22). To this end, we quantize the Hamiltonian
(26) by substituting the momentum operator,
$\hat{\bf p} = - i \hbar
\partial /\partial {\bf r}$, instead of the canonical momentum, ${\bf
p}$. For our problem, it is convenient to rewrite the obtained
quantized Hamiltonian in the following way:
\begin{equation}
\hat H = m_p c^2 + m_e c^2 + \frac{\hat {\bf
p}^2}{2m_e}-\frac{e^2}{r} + m_p \phi + \hat m^g_e \phi \ ,
\end{equation}
where electron passive gravitational mass operator is proportional
to its weight operator in the weak gravitational field (22),
\begin{equation}
\hat m^p_e  = m_e + \biggl(\frac{\hat {\bf p}^2}{2m_e}
-\frac{e^2}{r}\biggl)\frac{1}{c^2} + \biggl(2 \frac{\hat {\bf
p}^2}{2m_e}-\frac{e^2}{r} \biggl) \frac{1}{c^2} \ .
\end{equation}
It is important that the first term in Eq.(35) is the bare
electron mass, the second term is the expected electron energy
 contribution to the mass operator, whereas the
third nontrivial term is the virial contribution to the mass
operator. Note that, due to the existence of the virial operator
in Eq.(35), the electron passive gravitational mass operator does
not commute with electron energy operator, taken in the absence of
 the field.

\subsection{Hamiltonian approach [6,10]}
In this Subsection, we derive Hamiltonian (35) by means of two
methods. The first method, which we call semi-quantitative one,
uses the Schr\"{o}dinger equation in the curved spacetime of
General Relativity (22). The second method is related to direct
consideration of the Dirac equation in the curved spacetime (22).

\subsubsection{Semi-quantitative Hamiltonian [6]}
Let us consider a hydrogen atom near the Earth, where we use the
weak field approximation for the interval in gravitational field
(22). As mentioned in Subsection 4.1, we can introduce the local
proper spacetime coordinates (23), where the interval (22) has the
Minkowski form. In these local proper spacetime coordinates, the
Schr\"{o}dinger equation for electron in a hydrogen atom can be
approximately expressed in its standard form,
\begin{equation}
i \hbar \frac{\partial \Psi({\bf r'},t')}{\partial t'} = \hat H_0
(\hat {\bf p'},{\bf r'}) \ \Psi({\bf r'},t')  ,
\end{equation}
where
\begin{equation}
\hat H_0 (\hat {\bf p'},{\bf r'})= m_ec^2 + \frac{\hat {\bf
p'}^2}{2 m_e} - \frac{e^2}{r'} .
\end{equation}
We pay attention to the fact that Eqs.(36),(37) are written in the
so-called $1/c^2$ approximation. As to gravitational field, this
means that we take into account only terms of the order of
$|\phi|/c^2$. Near the Earth, this small parameter can be
estimated as $10^{-9}$, therefore, above we disregard terms of the
order of $(\phi/c^2)^2 \sim 10^{-18}$. We also point out that, as
usual, in Eqs.(36),(37), we do not take into account the so-called
tidal effects. This is equivalent to the fact that we do not
differentiate gravitational potential, $\phi$, with respect to
electron coordinates, ${\bf r}$ and ${\bf r'}$. Note that we also
use the approximation, $m_p \gg m_e$. Therefore, ${\bf r}$ and
${\bf r'}$, corresponding to electron positions in center of mass
coordinate system, we relate to its positions with respect to
proton. In the next Subsubsecttion, we show that, in fact,
ignoring all tidal effects means that we consider a hydrogen
atom as a point-like body. In particular, we disregard the tidal
terms in the electron Hamiltonian, which are very small and are of
the order of $(r_B/R_0)|\phi/c^2|(e^2/r_B) \sim
10^{-17}|\phi/c^2|(e^2/r_B)$ in the Earth's gravitational field.
We point out that, in Eqs.(36),(37), we also disregard magnetic
force and all spin related effects. Another our previously mentioned
suggestion is that proton mass is very high and, thus, proton can
be considered as a classical particle, whose position is fixed by
some non-gravitational force and whose kinetic energy is
negligible.

We also stress that, in this review, we consider the weak
gravitational field (22) as a perturbation in some inertial
coordinate system. The inertial coordinate system corresponds to
global spacetime coordinates, $(x,y,z,t)$ in Eq.(23), where it is
easy to obtain the following electron Hamiltonian from
Eqs.(36),(37):
\begin{equation}
\hat H(\hat {\bf p},{\bf r}) = m_e c^2 + \frac{\hat {\bf
p}^2}{2m_e}-\frac{e^2}{r} + m_e  \phi + \biggl( 3 \frac{\hat {\bf
p}^2}{2 m_e} -2\frac{e^2}{r} \biggl) \frac{\phi}{c^2}.
\end{equation}
It is important that the Hamiltonian (38) can be represented in
more convenient form,
\begin{equation}
\hat H(\hat {\bf p},{\bf r}) = m_e c^2 + \frac{\hat {\bf
p}^2}{2m_e} -\frac{e^2}{r} + \hat m^p_e \phi 
\ ,
\end{equation}
where we introduce the following electron passive gravitational
mass operator:
\begin{equation}
\hat m^p_e   = m_e  + \biggl(\frac{\hat {\bf
p}^2}{2m_e} -\frac{e^2}{r}\biggl) \frac{1}{c^2} + \biggl(2
\frac{\hat {\bf p}^2}{2m_e}-\frac{e^2}{r} \biggl) \frac{1}{c^2} ,
\end{equation}
which is proportional to electron weight operator in the weak
gravitational field (22). Note that, as usual, the gravitational mass
operator consists of three terms: the bare electron mass, $m_e$,
the expected electron energy contribution to the mass operator,
and the non-trivial virial contribution to passive gravitational
mass operator. It exactly coincides with electron passive gravitational
mass operator (35), obtained early by the Lagrangian method.

\subsubsection{More general Hamiltonian [18,10]}

The so-called gravitational Stark effect (i.e., the mixing effect
between even and odd wave functions in a hydrogen atom in gravitational field) was studied in Ref. [18] in the weak external gravitational field (22). Note that the corresponding Hamiltonian was derived in
$1/c^2$ approximation and a possibility of center of mass
of the atom motion was taken into account. The main peculiarity of the calculations in the above-mention paper was the fact that not only
terms of the order of $\phi/c^2$ were calculated, as in our case,
but also terms of the order of $\phi'/c^2$. Here, we use a
symbolic notation $\phi'$ for the first derivatives of
gravitational potential. In accordance with the existing
tradition, we refer to the latter terms as to the tidal ones. Note
that the Hamiltonian (3.24) was obtained in Ref. [18] directly from the
Dirac equation in a curved spacetime of General Relativity. As shown 
in [18], it can be rewritten for the corresponding Schr\"{o}dinger equation as a sum of the four terms:
\begin{equation}
\hat H (\hat {\bf P}, \hat {\bf p}, \tilde {\bf R},r)= \hat H_0
(\hat {\bf P}, \hat {\bf p}, r) + \hat H_1 (\hat {\bf P}, \hat
{\bf p}, \tilde {\bf R},r) + \hat H_2 (\hat {\bf p}, {\bf r}) +
\hat H_3 (\hat {\bf P}, \hat {\bf p},\tilde {\bf R},r) ,
\end{equation}

\begin{equation}
\hat H_0 (\hat {\bf P}, \hat {\bf p}, r) = m_e c^2 + m_p c^2 +
\biggl[\frac{\hat {\bf P}^2}{2(m_e + m_p)} + \frac{\hat {\bf
p}^2}{2 \mu} \biggl] - \frac{e^2}{r} ,
\end{equation}

\begin{equation}
\hat H_1 (\hat {\bf P}, \hat {\bf p}, \tilde {\bf R}, r) =
\biggl\{ m_e c^2 + m_p c^2 +  \biggl[3 \frac{\hat {\bf P}^2}{2(m_e
+ m_p)} + 3 \frac{\hat {\bf p}^2}{2 \mu} - 2 \frac{e^2}{r}
\biggl]\biggl\} \biggl( \frac{\phi  - {\bf g}\tilde {\bf R}}{c^2}
\biggl),
\end{equation}

\begin{equation}
\hat H_2 (\hat {\bf p}, {\bf r}) = \frac{1}{c^2}
\biggl(\frac{1}{m_e}-\frac{1}{m_p} \biggl)[-({\bf g}{\bf r}) \hat
{\bf p}^2 + i \hbar {\bf g} \hat {\bf p}] + \frac{1}{c^2} {\bf g}
\biggl(\frac{\hat {\bf s_e}}{m_e} - \frac{\hat {\bf s_p}}{m_p}
\biggl) \times \hat {\bf p} + \frac{e^2 (m_p-m_e)}{2(m_e+m_p)c^2}
\frac{{\bf g}{\bf r}}{r},
\end{equation}

\begin{equation}
\hat H_3 (\hat {\bf P}, \hat {\bf p}, \tilde {\bf R}, r) =
\frac{3}{2}\frac{i \hbar {\bf g}{\bf P}}{(m_e+m_p)c^2}
+\frac{3}{2} \frac{{\bf g}{\bf(s_e+s_p)}\times {\bf
P}}{(m_e+m_p)c^2} - \frac{({\bf g}{\bf r})({\bf P}{\bf p})+({\bf
P}{\bf r})({\bf g}{\bf p})-i\hbar {\bf g}{\bf P}}{(m_e+m_p)c^2},
\end{equation}
where ${\bf g}=-G \frac{M}{R^3} {\bf R}$. Note that we use the
following notations in Eqs.(41)-(45): $\tilde {\bf R}$ and ${\bf
P}$ stand for coordinate and momentum of a hydrogen atom center of
mass, respectively; whereas, ${\bf r}$ and ${\bf p}$ stand for
relative electron coordinate and momentum in center of mass
coordinate system; $\mu = m_e m_p /(m_e + m_p)$ is the so-called
reduced electron mass. We point out that $\hat H_0 (\hat {\bf P},
\hat {\bf p}, r)$ is the Hamiltonian of a hydrogen atom in the
absence of the field. It is important that the Hamiltonian $\hat
H_1 (\hat {\bf P}, \hat {\bf p}, \tilde {\bf R}, r)$ describes
couplings not only of the bare electron and proton masses with the
gravitational field (22) but also couplings of electron kinetic
and potential energies with the field. And finally, the
Hamiltonians $\hat H_2 (\hat {\bf p}, {\bf r})$ and $\hat H_3
(\hat {\bf P}, \hat {\bf p}, \tilde {\bf R}, r)$ describe only the
tidal effects.

Let us strictly derive the Hamiltonian (39),(40), which has been
semi-quantitatively derived in Subsubsection 4.2.1, from the more
general Hamiltonian (41)-(45). As was already mentioned, we use
the approximation, where $m_p \gg m_e$, and, therefore, $\mu =
m_e$. In particular, this allows us to consider proton as a heavy
classical particle. We recall that we need to derive the
Hamiltonian of the atom, whose center of mass is at rest with
respect to the Earth. Thus, we can omit center of mass kinetic
energy and center of mass momentum. As a result, the first two
contributions to electron part of the total Hamiltonian (41)-(45)
can be written in the following way:
\begin{equation}
\hat H_0 (\hat {\bf p}, r) = m_e c^2 + \frac{\hat {\bf p}^2}{2m_e}
- \frac{e^2}{r}
\end{equation}
and
\begin{equation}
\hat H_1 (\hat {\bf p}, r) =  \biggl\{ m_e c^2 + \biggl[3
\frac{\hat {\bf p}^2}{2 m_e} - 2 \frac{e^2}{r} \biggl]\biggl\}
\biggl( \frac{\phi}{c^2} \biggl),
\end{equation}
where we place center of mass of the atom at point $\tilde {\bf R}
= 0$. Now, let us study the first tidal term (44) in the total
Hamiltonian (41). At first, we pay attention that $|{\bf g}|
\simeq |\phi|/R_0$. Then, as well known, in a hydrogen atom
$|{\bf r}| \sim \hbar / |{\bf p}| \sim r_B$ and ${\bf p}^2/(2m_e)
\sim e^2/r_B$. These values allow us to evaluate the first tidal
term (44) in the Hamiltonian (41) as $H_2 \sim (r_B/R_0)
(|\phi|/c^2) (e^2/r_B) \sim 10^{-17} (|\phi|/c^2) (e^2/r_B)$. Note
that this value is $10^{-17}$ times smaller than $H_1 \sim
(|\phi|/c^2) (e^2/r_B)$ and $10^{-8}$ times smaller than the
second correction with respect to the small parameter
$|\phi|/c^2$. Therefore, we can disregard the
contribution (44) to the total Hamiltonian (41). As to the
second tidal term (45) in the total Hamiltonian, we pay attention
that it is exactly zero in the case, where ${\bf P}=0$, considered
in this review. Therefore, we can conclude that the Hamiltonian
(46),(47), derived in this Subsubsection, exactly coincides with
that, semi-quantitatively derived by us in Refs.[4-7] [see
Eqs.(39),(40)].

\subsection{Equivalence of the expectation values [4-7]}

In this Subsection, we obtain an important consequence of
Eqs.(39),(40). Note that the electron passive gravitational mass
operator (40) does not commute with the electron energy operator,
taken in the absence of the gravitational field. Thus, it seems
that there is no any equivalence between electron passive
gravitational mass and its energy. But this is not true and below
we establish the equivalence between electron energy and the
expectation value of electron passive gravitational mass for
stationary quantum states. To show their equivalence, we consider
a macroscopic ensemble of hydrogen atoms with each of them being
in n-th stationary state with energy $E_n$,
\begin{equation}
\Psi_n(r,t) = \Psi_n(r) \exp \biggl( \frac{-im_ec^2t}{\hbar}
\biggl) \exp \biggl( \frac{- iE_nt}{\hbar} \biggl)\ ,
\end{equation}
where $\Psi_n(r)$ is a normalized electron wave
function of n-th energy level in a hydrogen atom.
From Eq.(40), it follows that the expectation value of electron
passive gravitational mass operator in this case is
\begin{equation}
<\hat m^p_e > = m_e + \frac{ E_n}{c^2}  + \biggl< 2 \frac{\hat
{\bf p}^2}{2m_e}-\frac{e^2}{r} \biggl> \frac{1}{c^2} = m_e +
\frac{E_n}{c^2} .
\end{equation}
Here, as it was for active gravitational mass operator (14), the
expectation value of the virial term in Eq.(49) is zero due to the
quantum virial theorem [17]. Therefore, we conclude that the
equivalence between passive gravitational mass and energy exists
at a macroscopic level for stationary quantum states [4-7,10].

\subsection{Inequivalence between passive gravitational mass
and energy at a macroscopic level [4-7,10]}

In the previous Subsection, we demonstrated that energy was
equivalent to the expectation value of passive gravitational mass
for stationary quantum states. Below, we make the following
statement. We stress that, for superposition of stationary quantum
states, the expectation value of passive gravitational mass can be
oscillatory function of time even in the case, where the
expectation value of energy is constant. Here, as in the case of
active gravitational mass, we consider the simplest superposition
of 1S and 2S energy levels (17),
\begin{equation}
\Psi_{1,2}(r,t) = \frac{1}{\sqrt{2}} \bigl[ \Psi_1(r) \exp(-iE_1t)
+ \Psi_2(r) \exp(-iE_2t) \bigl],
\end{equation}
which is characterized by the time-independent expectation value of
energy,
\begin{equation}
<E> = \frac{E_1+E_2}{2} .
\end{equation}
By using Eq.(40), it is easy to show that, for the wave function 
(50), the expectation value of electron
 passive gravitational mass is the following oscillatory function:
\begin{equation}
<\hat m^p_e> = m_e + \frac{E_1+E_2}{2 c^2} + \frac{V_{1,2}}{c^2}
\cos \biggl[ \frac{(E_1-E_2)t}{\hbar} \biggl] ,
\end{equation}
where matrix element of the virial operator, $V_{1,2}$ is defined
by Eq.(20). In our opinion, the time-dependent oscillations of the
passive gravitation mass (52) directly demonstrate breakdown of
the equivalence between passive gravitational mass and energy at a
macroscopic level. It is important that these oscillations are of
the order of $\alpha^2 m_e$ (i.e. they are strong enough) and are
of a pure quantum origin without classical analogs, where $\alpha$
is the fine structure constant. We also pay attention that the
similar oscillations exist for active gravitational mass of
quantum superposition of stationary states [see Eq. (19)]. We hope
that these strong oscillations of passive and active gravitational
masses are experimentally measured, despite the fact that the
quantum states (17),(50) decay with time.

If we average the oscillations (52) over time, we obtain the
modified equivalence principle between the averaged over time
expectation value of passive gravitational mass and the
expectation value of energy in the following form:
\begin{equation}
<< \hat m^p_e >>_t = m_e + \frac{E_1+E_2}{2c^2} = \frac{<E>}{c^2}.
\end{equation}
We stress that physical meaning of averaging procedure in Eq.(53)
is completely different from that of classical time averaging
procedure (27) and does not have the corresponding classical
analogs.

\subsection{Inequivalence between passive gravitational mass
and energy at a microscopic level [4-7]}

Here, we describe Gedanken experiment, which directly
demonstrates breakdown of the equivalence between passive
gravitational mass and energy at a microscopic level.
At first, we consider electron in its ground state  in a hydrogen
 atom with the following wave function, corresponding
to the absence of the gravitational field (22),
\begin{equation}
\Psi_1(r,t) = \Psi_1(r) \exp \biggl( \frac{-im_ec^2t}{\hbar}
\biggl) \exp \biggl( \frac{- iE_1t}{\hbar} \biggl)\ ,
\end{equation}
where
\begin{equation}
\hat H_0(\hat {\bf p},r) \Psi_1(r) = E_1 \Psi_1(r) , \ \ \ \hat
H_0(\hat {\bf p},r)= m_ec^2 + \frac{\hat {\bf p}^2}{2 m_e}-
\frac{e^2}{r}.
\end{equation}
Now, we account for the gravitational field (22), as a
perturbation to the Hamiltonian (55),
\begin{equation}
\hat H(\hat {\bf p},{\bf r}) = \hat H_0(\hat {\bf p},r) + \hat m^p_e \phi \ ,
\end{equation}
where electron passive gravitational mass operator is defined by
Eq.(40). Ground state wave function of the Hamiltonian (56),
$\tilde \Psi_1(r)$, in accordance with the standard quantum
mechanical perturbation theory, can be written as
\begin{equation}
\tilde \Psi_1(r) = \sum_n a_n \Psi_n(r) ,
\end{equation}
where
\begin{equation}
\hat H(\hat {\bf p},{\bf r}) \tilde \Psi_1(r) = \tilde E_1 \tilde
\Psi_1(r).
\end{equation}

We pay attention that, due to selection rules of the passive
 gravitational mass operator (40), $\Psi_n (r)$ are normalized 
 electron wave functions in the absence of the gravitation
(22), corresponding only to atomic levels $nS$ with energy $E_n$. 
Let us define coefficient $a_1$ and correction to energy of the 
ground state. In accordance with the perturbation theory, they can
be written as:
\begin{equation}
a_1 \simeq 1, \ \ \ \tilde E_1 = \biggl(1 + \frac{\phi}{c^2}
\biggl) E_1 .
\end{equation}
Here, the last term in Eq.(59) represents the famous red shift
in the gravitational field (22). Note that it is the expected contribution to passive gravitational mass due to electron binding
energy in the atom. We pay attention that to derive Eq.(59), we have
used the quantum virial theorem [17] in the following form:
\begin{equation}
\int \Psi_1(r) \biggl( 2 \frac{\hat {\bf p}^2}{2 m_e} -
\frac{e^2}{r} \biggl) \Psi_1(r) d^3 {\bf r} = 0 .
\end{equation}
On the other hand, the coefficients $a_n$ with $n \neq 1$ in Eq.(58) can be expressed through the matrix elements of the virial operator,
\begin{equation}
a_n = \biggl( \frac{\phi}{c^2} \biggl) \biggl(
\frac{V_{n,1}}{E_1-E_2} \biggl), \ \ \ V_{n,1} = \int \Psi_n(r)
\biggl( 2 \frac{\hat {\bf p}^2}{2 m_e} - \frac{e^2}{r} \biggl)
\Psi_1(r) d^3 {\bf r} .
\end{equation}
It is important that the obtained electron wave function (57)-(61),
 which corresponds to ground state in the presence of the gravitational
field (22), is written as a series of eigenfunctions of electron energy
operator in the absence of the field. Thus, if we would like to measure
energy by means of operator (55), we will obtain the following
 quantized values:
\begin{equation}
E(n) = m_e c^2 + E_n .
\end{equation}
Therefore, we conclude that the Einstein's equation, $E = m_e c^2 + E_1$, is broken in our case with small but finite probabilities [4-7],
\begin{equation}
P_n = |a_n|^2 = \biggl(\frac{\phi}{c^2}\biggl)^2
\frac{V^2_{n,1}}{(E_n -E_1)^2} , \ \ \ n \neq 1 .
\end{equation}
Note that the reason for this breakdown of the equivalence between
passive gravitational mass and energy is that electron wave function
 with definite passive gravitational mass (57)-(61) is not characterized by definite energy in the absence of the gravitational field (22). 

\section{Summary}
In conclusion, in the review, we have discussed in detail breakdown of
 the equivalence between active and passive gravitational masses of
 an electron and its energy in a hydrogen atom. We stress that the considered phenomena are very general and are not restricted by atomic physics and the Earth's gravitational field. In other words, the above discussed phenomena exist for any quantum system and any gravitational field.  In this review, we also have improved several drawbacks of the original pioneering works.

\section*{Acknowledgments}

We are thankful to N.N. Bagmet (Lebed), V.A. Belinski, Steven
Carlip, Fulvio Melia, Jan Rafelski, Douglas Singleton, Elias Vagenas, and V.E. Zakharov for fruitful and useful discussions.

\end{document}